\documentclass[a4paper, 12pt]{article}
\usepackage{xcolor,graphicx}

\usepackage[pdftex, urlcolor=black]{hyperref}

\title{\textsf{qrypt0} - encrypted short messages\\ exchanged between offline computers}
\author{Andreas O. Bender\\
\href{http://www.andreasobender.org}{\texttt{www.andreasobender.org}}\\
\href{mailto:andreasobender@mac.com}{\texttt{andreasobender@mac.com}}}

\begin{document}
\maketitle

\begin{abstract}
A system is described for exchanging encrypted short messages between computers which remain permanently isolated from any network accessible to the attacker. The main advantage is effective protection of these computers from malware which could circumvent the encryption. For transmission, the ciphertext is passed between isolated and connected computers in the form of a QR code\footnote{"QR Code" is a registered trademark of DENSO WAVE INCORPORATED.}, which is displayed on and scanned from a screen. The security of \textsf{qrypt0} therefore rests on the cryptography and the computer's physical isolation rather than on the computer security of the encrypting device. 
\end{abstract}

\section{\textsf{crypto} on networked devices}

Message encryption is usually carried out on networked computers. This is hardly every questioned as a means to transmit the ciphertext from Alice to Bob, even though it allows an attack which is not directed at the cryptography: It is directed at the endpoints which carry out the encryption. 

This is highly relevant in view of the fact that rather than attacking the encryption, circumventing it is often much easier. Six possibilities for doing so are listed in~\cite{kerrSchneier}. Four of them consist of guessing, finding or compelling the key and locating the plaintext. Defending against these circumventions is possible by the corresponding measures of choosing a key of sufficient entropy, protecting the key and keeping track of all copies of the plaintext. 

The plaintext must of course be stored on the encrypting device for at least some time. Therefore the other two possibilities of circumventing encryption are hard to defend against: exploiting a flaw in the encrypting device and accessing the plaintext when in use. In practice attackers use vulnerabilities of the operating system for things like the installation of keyloggers which record the plaintext as it is typed in. 

Some known examples of such attacks are recently disclosed activities in the program Vault 7 by the United States' Central Intelligence Agency~\cite{vault7}. The German Federal Intelligence Service BND has started a program to circumvent encryption in various popular messengers~\cite{bnd}. 

Vulnerabilities which are not publicly known, so-called Zero-Days, are traded continuously by companies like Zerodium~\cite{zerodium}. Security patches are released for the gaps which do become public knowledge. However, the author is not aware of any producers of system software who are even prepared to suggest, let alone give an assurance that there are no more such vulnerabilities hidden in their software. 

Operating systems impervious to infection by malware appear to be a long way off~\cite[chpt. 8]{ferSchnKoh}~\cite[20.6]{bishop}. 

\bigskip
We need  a system with the following properties:
\begin{itemize}
\item The exchange of encrypted messages can be protected effectively against circumvention of the encryption. In particular, security vulnerabilities in the operating system of the encrypting computer do not compromise the system's security. 
\item The attacker sees only the number of messages, their timing and the ciphertext. 
\end{itemize} 
Relying only on currently available technology, compromises regarding convenience are to be expected. Even so, for high-value data a system offering this level of security with some inconvenience is still much better than what has been available up to now. 

\newpage
\section{\textsf{qrypt0} on isolated devices}

The distinguishing feature of \textsf{qrypt0} is the isolation of the encrypting computer from any open network  and the transmission of the ciphertext encoded in a QR code. This barcode is displayed on Alice's screen and scanned into a connected computer which sends it on to Bob. 

This isolation from the network provides effective protection against remote infection with malware. Keeping the isolated computer physically secure is necessary to protect it from local infection with malware. 

The website~\cite{qrypt0} links to a freely available prototype, which includes a discussion of implementation issues. 

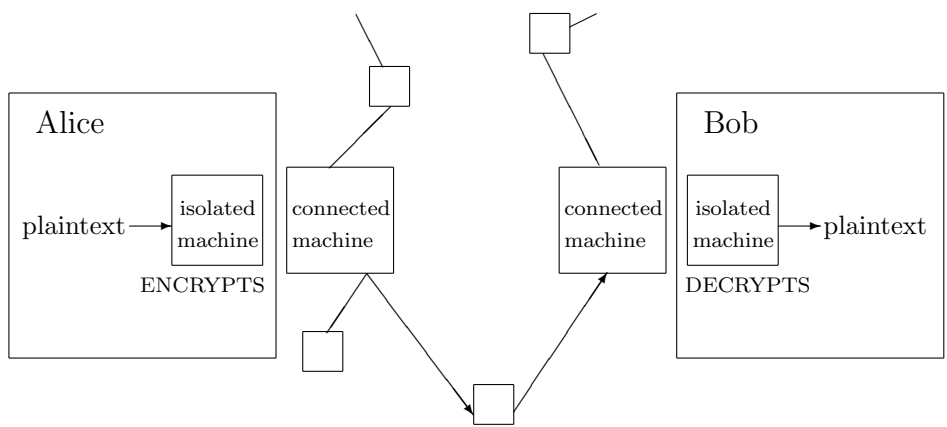
\begin{picture}(350, 180)(0,0)
\newsavebox{\bigbox}
\savebox{\bigbox}
(100,100)[bl]{
\put(0,0){\line(1,0){100}}
\put(100,0){\line(0,1){100}}
\put(0,0){\line(0,1){100}}
\put(0,100){\line(1,0){100}}
}
\newsavebox{\mediumbox}
\savebox{\mediumbox}
(40,40)[bl]{
\put(0,0){\line(1,0){40}}
\put(40,0){\line(0,1){40}}
\put(0,0){\line(0,1){40}}
\put(0,40){\line(1,0){40}}
}
\newsavebox{\smallbox}
\savebox{\smallbox}
(34,34)[bl]{
\put(0,0){\line(1,0){34}}
\put(34,0){\line(0,1){34}}
\put(0,0){\line(0,1){34}}
\put(0,34){\line(1,0){34}}
}
\newsavebox{\tinybox}
\savebox{\tinybox}
(15,15)[bl]{
\put(0,0){\line(1,0){15}}
\put(15,0){\line(0,1){15}}
\put(0,0){\line(0,1){15}}
\put(0,15){\line(1,0){15}}
}

\put(0,25){\usebox{\bigbox}}
\put(61,60){\usebox{\smallbox}}
\put(10,110){Alice}
\put(5,72){\footnotesize plaintext}
\put(64,79){\scriptsize isolated}
\put(63,67){\scriptsize machine}
\put(49,50){\scriptsize \textsc ENCRYPTS}
\put(45,75){\vector(1,0){16}}

\put(104,57){\usebox{\mediumbox}}
\put(106,79){\scriptsize connected}
\put(106,67){\scriptsize machine}
\put(134,57){\vector(3,-4){40}}

\put(110,20){\usebox{\tinybox}}
\put(134,57){\line(-2,-3){15}}

\put(135,120){\usebox{\tinybox}}
\put(120,97){\line(1,1){23}}

\put(140,135){\line(-1,2){10}}

\put(195,140){\usebox{\tinybox}}
\put(200,140){\line(1,-2){21}}
\put(210,150){\line(2,1){10}}

\put(174,0){\usebox{\tinybox}}
\put(189,5){\vector(2,3){35}}

\put(206,57){\usebox{\mediumbox}}
\put(208,79){\scriptsize connected}
\put(208,67){\scriptsize machine}

\put(250,25){\usebox{\bigbox}}
\put(254,60){\usebox{\smallbox}}
\put(260,110){Bob}
\put(305,72){\footnotesize plaintext}
\put(257,79){\scriptsize isolated}
\put(256,67){\scriptsize machine}
\put(253,50){\scriptsize \textsc DECRYPTS}
\put(288,75){\vector(1,0){16}}

\end{picture}

\subsection{Functionality}

{\bf Message encryption.} The message is entered into the isolated computer by Alice. The system adds a message number and a timestamp. The message is authenticated, padded up to the maximum length for one barcode and encrypted. The resulting ciphertext is encoded in a QR code.  

\smallskip
\noindent {\bf Message transmission.} An online device under Alice's control scans in the QR code from the screen of the isolated computer and transmits the barcode itself or the ciphertext to the recipient by any means available. 

\smallskip
\noindent {\bf Message decryption.} An online device under Bob's control receives the message containing the ciphertext. If necessary, this device encodes the ciphertext into a QR code and displays it on screen. Bob's isolated computer scans in the QR code, decodes and decrypts the message, removes the pad and checks the authentication. 

\subsection{Design choices}

The main cost of the system consists of one additional computer per user and a limit on the length of the message which one QR code can contain. The QR code has the largest capacity of 2953 bytes among common 2-dimensional barcodes. Note that this includes the characters needed for formatting purposes, for authentication, the message number and a timestamp. 

\smallskip
{\bf Installation.} In order to avoid infection with malware, the necessary software should be installed by burning it onto an optical disk, transferring it to the isolated machine from that medium and then checking its authentication. 

\smallskip
{\bf Cryptography.} The system can be used with both asymmetric and symmetric cryptography. Using only symmetric cryptography requires a pre-shared key, but has the advantage of simplicity. No understanding of handling public and private keys is required. 

If asymmetric cryptography is used, key material can be entered by scanning QR codes, where more than one may be required. 

\smallskip
{\bf Formatting.} In order to be sure no messages got lost, a message number should be added. Management of messages is easier to organize if also a timestamp is included. 

\smallskip
{\bf Authentication.} This can  be done using a symmetric key for a hash-based message authentication code. The choice of whether encryption or authentication is done first has to be made. 

\smallskip
{\bf Message length.} We recommend padding the message up to its maximal length, since information about the message length may be valuable to an attacker. 

\smallskip
Note that no information on the message is transmitted by the system except the ciphertext itself. In particular, the file name has to be chosen anew after passing the ciphertext between two computers by scanning in a QR code. 

\smallskip
{\bf System use information.} Either the QR code itself or the ciphertext it contains can be transmitted. Transmitting the QR code itself passes information on system use to an attacker. 

\smallskip
{\bf Plaintext storage.} As mentioned above, one way of circumventing encryption consists of locating copies of the plaintext. As protection against this, we recommend to do a secure erase of the plaintext messages after display and to retain only the ciphertext on secondary storage.

\newpage
\bigskip
{\bf System advantages:}
\begin{itemize}
\item The only way the system can be compromised by malware after installation is by gaining physical access to the isolated computer. 
\item Sidechannel attacks using timing information or measurement of power consumption are only possible with physical access to the isolated computer while the system is running. 
\item In case only symmetric cryptography is used, the system is easy to handle. The key consists of a sufficiently long passphrase. Neither management of keys on secondary storage nor an understanding of asymmetric cryptography is necessary.   
\end{itemize}

\smallskip
{\bf System limitations:}
\begin{itemize}
\item One QR code can contain at most 2953 bytes, including some bytes used for formatting. Depending on the scanning device used, the limit can be about half that number. 
\item Every user needs an isolated computer. 
\item The isolated computer needs to be kept physically secure. 
\end{itemize}

\smallskip
{\bf Attacks:}
\begin{itemize}
\item Evil maid attack. Anyone who has physical access to the isolated computer can install malware. Such a program can establish and use a hidden channel in the QR code or other sidechannels like the loudspeaker or flashing lights. 
\item RF emissions of the isolated machine as side channel. 
\item Traffic analysis. The attacker can see the number and timing of transmitted messages. 
\item Ciphertext-only attack on the cipher itself. 
\end{itemize}


\newpage
\begin{thebibliography}{11}

\bibitem{kerrSchneier}
\textsc{Orin S. Kerr, Bruce Schneier.} Encryption workarounds. 
\verb+https://papers.ssrn.com/sol3/papers.cfm?abstract_id=2938033+

\bibitem{vault7}
\verb+https://wikileaks.org/ciav7p1/+ Released March 7, 2017.

\bibitem{bnd}
\verb+https://netzpolitik.org/2016/projekt-aniski-wie-der-bnd-mit-+
\verb+150-millionen-euro-messenger-wie-whatsapp-entschluesseln-will/+

\bibitem{zerodium}
\verb+http://www.zerodium.com+

\bibitem{ferSchnKoh}
\textsc{Niels Ferguson, Bruce Schneier, Tadayoshi Kohno.} Cryptography engineering. \emph{Wiley Publishing 2010.} 

\bibitem{bishop}
\textsc{Matt Bishop.} Computer security: art and science. \emph{Addison-Wesley 2002.}

\bibitem{qrypt0}
\verb+www.andreasobender.org+

\end{thebibliography}
\end{document}